\documentclass[english,prb,twocolumn,showpacs,
superscriptaddress,floatfix,longbibliography]{revtex4-1}

\usepackage[T1]{fontenc}
\usepackage[latin9]{inputenc}
\usepackage{amsmath,amssymb,amstext}
\usepackage{xcolor}
\usepackage{graphicx}
\usepackage{esint}
\usepackage{babel}
\usepackage{verbatim}
\usepackage{color}
\usepackage{subfigure}
\usepackage{braket}
\usepackage[pdfborder={0 0 1}]{hyperref}
\definecolor{TODO-color}{named}{blue}

\definecolor{SC-color}{named}{red}

\definecolor{FA-color}{named}{cyan}

\definecolor{TTH-color}{named}{green}

\definecolor{TTH-color}{named}{green}

\definecolor{SB-color}{named}{blue}

\definecolor{MR-color}{rgb}{0.11, 0.55, 0.02}

\begin{document}

\title{Charge transport through spin-polarized tunnel junction between two
  spin-split superconductors }

\author{Mikel Rouco} \email{mikel.rouco@ehu.eus} \affiliation{Centro de F\'isica
  de Materiales (CFM-MPC), Centro Mixto CSIC-UPV/EHU, Manuel de Lardizabal 5,
  E-20018 San Sebasti\'an, Spain}

\author{Subrata Chakraborty} \email{schrkmv@gmail.com} \affiliation{Department
  of Physics and Nanoscience Center, University of Jyv{\"a}skyl{\"a}, P.O. Box
  35 (YFL), FI-40014 Jyv{\"a}skyl{\"a}, Finland}\affiliation{Department of
  Physics, Queens College of the City University of New York, Queens, NY 11367,
  USA}

\author{Faluke Aikebaier} \affiliation{Department of Physics and Nanoscience
  Center, University of Jyv{\"a}skyl{\"a}, P.O. Box 35 (YFL), FI-40014
  University of Jyv{\"a}skyl{\"a}, Finland}
 
\author{V. N. Golovach} \affiliation{Centro de F\'isica de Materiales (CFM-MPC),
  Centro Mixto CSIC-UPV/EHU, Manuel de Lardizabal 5, E-20018 San Sebasti\'an,
  Spain} \affiliation{IKERBASQUE, Basque Foundation for Science, 48013 Bilbao,
  Basque Country, Spain} \affiliation{Donostia International Physics Center
  (DIPC),Manuel de Lardizabal 4, E-20018 San Sebastian, Spain}
 
\author{Elia Strambini} \affiliation{NEST Istituto Nanoscienze-CNR and Scuola
  Normale Superiore, I-56127 Pisa, Italy}

\author{Jagadeesh S. Moodera} \affiliation{Department of Physics and Francis
  Bitter Magnet Lab, Massachusetts Institute of Technology, Cambridge,
  Massachusetts 02139, USA}

\author{Francesco Giazotto} \email{francesco.giazotto@sns.it} \affiliation{NEST
  Istituto Nanoscienze-CNR and Scuola Normale Superiore, I-56127 Pisa, Italy}

\author{Tero T. Heikkil{\"a}} \email{tero.t.heikkila@jyu.fi}
\affiliation{Department of Physics and Nanoscience Center, University of
  Jyv{\"a}skyl{\"a}, P.O. Box 35 (YFL), FI-40014 University of
  Jyv{\"a}skyl{\"a}, Finland}

\author{F. Sebastian Bergeret} \email{fs.bergeret@csic.es} \affiliation{Centro
  de F\'isica de Materiales (CFM-MPC), Centro Mixto CSIC-UPV/EHU, Manuel de
  Lardizabal 5, E-20018 San Sebasti\'an, Spain} \affiliation{Donostia
  International Physics Center (DIPC),Manuel de Lardizabal 4, E-20018 San
  Sebastian, Spain}

\date{\today}

\begin{abstract}
  We investigate transport properties of junctions between two spin-split
  superconductors linked by a spin-polarized tunneling barrier. The
  spin-splitting fields in the superconductors (S) are induced by adjacent
  ferromagnetic insulating (FI) layers with arbitrary magnetization. The aim of
  this study is twofold: On the one hand, we present a theoretical framework
  based on the quasiclassical Green's functions to calculate the Josephson and
  quasiparticle current through the junctions in terms of the different
  parameters characterizing it. Our theory predicts qualitative new results for
  the tunneling differential conductance, $dI/dV$, when the spin-splitting
  fields of the two superconductors are non-collinear. We also discuss how
  junctions based on FI/S can be used to realize anomalous Josephson junctions
  with a constant geometric phase shift in the current-phase relation. As a
  result, they may exhibit spontaneous triplet supercurrents in the absence of a
  phase difference between the S electrodes. On the other hand, we show results
  of planar tunneling spectroscopy of a EuS/Al/AlO$_x$/EuS/Al junction and use
  our theoretical model to reproduce the obtained $dI/dV$ curves.  Comparison
  between theory and experiment reveals information about the intrinsic
  parameters of the junction, such as the size of the superconducting order
  parameter, spin-splitting fields and spin relaxation, and also about
  properties of the two EuS films, such as their morphology, domain structure,
  and magnetic anisotropy.
\end{abstract}

\maketitle

\section{Introduction}
\label{sec:introduction}

Superconducting films with spin-split density of states have been used for a
long time to determine the spin polarization of ferromagnetic metals
tunnel-coupled to the superconductor (S)
\cite{PhysRevLett.25.1270,meservey1975tunneling,PhysRevLett.26.192,tedrow1973spin,PhysRevB.16.4907,
  PhysRevB.22.1331,meservey1994spin}.  Originally, the spin splitting was
induced by applying in-plane magnetic fields to thin superconducting
films. These fields had to be large, of the order of few Tesla, in order to
obtain sizable splittings. Interestingly, as shown in the late 1980s, such spin
splitting can also be observed at rather small, or even zero, magnetic fields in
superconducting Al layers adjacent to ferromagnetic insulators
(FI)\cite{moodera-electron-1988,Hao:1990}. In this case the
splitting is attributed to the exchange interaction at the FI/S
interface\cite{PhysRevB.38.8823}. Additionally, those first works on FI/S
structures showed that thin FI layers can also be used as very efficient
spin-filters, with potential application as sources for highly spin-polarized
spin currents\cite{Moodera_review}.

More recently, non-equilibrium properties of superconductors with a spin-split
density of states have attracted a renewed attention
\cite{hubler2010charge,Wolf2014,beckmann2016spin,Kolenda2016,silaev2015long,bergeret2018colloquium,heikkila2019thermal}. In
such systems, two additional spin-dependent modes appear and couple to the
widely studied non-equilibrium energy and charge modes
\cite{silaev2015long,schmid1975linearized}.  FI/S structures have also been
suggested for several applications, as highly efficient thermoelectric
elements\cite{MacHon2013,Ozaeta2014a},
bolometers\cite{heikkila2018thermoelectric}, thermometers\cite{giazotto2015b},
cryogenic RAM memories\cite{de2018toward}, and different caloritronic devices to
access the electronic heat current in nanostructures
\cite{Giazotto:2013ei,Giazotto2014a,giazotto2015,rouco2018electron,chakraborty}.

Most of these applications require both superconductors with spin-split density
of states and highly polarized spin-filter interfaces. This motivates the
present work, in which we explore both theoretically and experimentally FI/S
junctions.  Theoretically, we develop a general model to describe the coupling
of two spin-split superconductors through an additional spin-filter barrier. Our
model takes into account self-consistently magnetic disorder, spin-orbit
coupling, and orbital effects of the magnetic field, as well as non-collinear
spin-splitting fields. On the one hand our model predicts new features in
FI/S-based junctions: additional coherent peaks in the differential conductance
when the FI layers are monodomain with non-collinear magnetization, and the
possible realization of an anomalous Josephson junction with pure triplet
supercurrents at zero phase bias.  On the other hand, our model provides a tool
to interpret transport experiments on tunneling junctions with FI/S electrodes.
 
Experimentally, we measure the tunneling conductance of an EuS/Al/AlO$_x$/EuS/Al
junction as a function of the applied voltage and magnetic field. The
differential tunneling conductance, $dI/dV$, shows sharp peaks whose heights
depend on the effective spin splitting induced in both Al layers and the spin
filtering of the barrier. Below, we perform a self-consistent calculation that
allows us to determine unambiguously the main parameters governing the transport
of the junction

The work is organized as follows. In the next section we present the
measurements of the tunneling conductance of the junction under consideration as
a function of the magnetic field. In Sec.~\ref{sec:model} we present a
theoretical model based on the quasiclassical Green's functions for the
description of the transport properties of a generic FI/S/I/FI/I/S/FI
junction. In Sec.~\ref{sec:Josephson} we discuss the Josephson current through
such junctions with emphasis on the anomalous behavior when the FI
magnetizations are non-collinear. In Sec.~\ref{sec:discussion} we focus on the
quasiparticle current and the tunneling differential conductance. The latter is
compared to the experimental data, and a discussion of the results follows. We
present our conclusions in Sec.~\ref{sec:conclusions}.

\section{Tunneling Conductance of a EuS/Al/AlO$_x$/EuS/Al junction }
\label{sec:experiment}

\begin{figure}[t!]
  \centering \includegraphics[width=\linewidth]{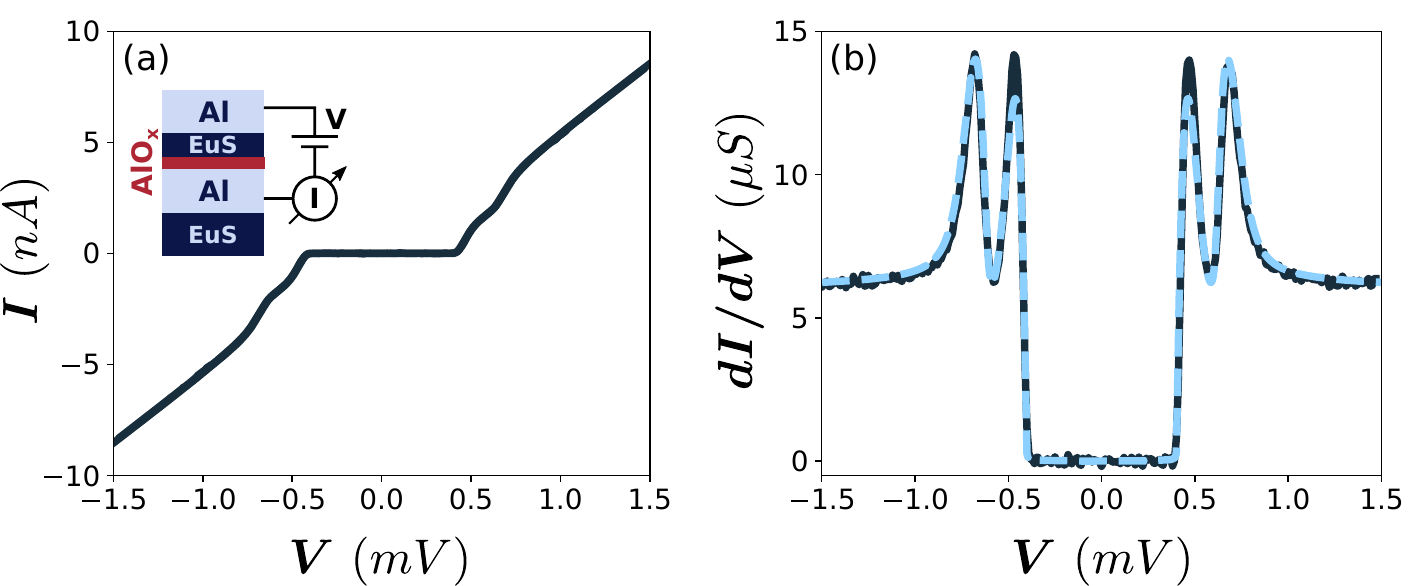}
  \caption{Tunneling spectroscopy of a FI/S/I/FI/S junction before applying an
    external magnetic field. (a) Typical current ($I$) vs voltage ($V$)
    characteristic of the junction measured at 25 mK. (b) Numerical derivative
    of the $I-V$ characteristic extracted from the data in panel (a)(black
    line). The blue dashed line is obtained from our theoretical model presented
    in Sec.~\ref{sec:model}. The parameters used for the fitting are: $G_T=6$
    $\mu$S, $\Delta_0=320$ $\mu$eV, $h_L = 0$, $h_R = 100$ $\mu$eV,
    $\tau_{sf}^{-1} = 0.08\Delta_0$ and $\tau_{so}^{-1} = \tau_{orb}^{-1} =
    0$. In the demagnetized regime, the effective spin splitting in the upper Al
    layer is negligibly small. The spin splitting arises from the very large
    domain structure of the bottom EuS layer, with size much larger than the
    superconducting coherence length $\xi_0$. The measured peak structure
    resembles the one measured in Ref.~\cite{strambini-2017-revealing} without
    the spin-filtering effect at work [see discussion after
    Eq.~\eqref{eq:current-demagnetized} for more details]. }
  \label{fig:Exp1}
\end{figure}
\begin{figure}[t!]
  \centering \includegraphics[width=\linewidth]{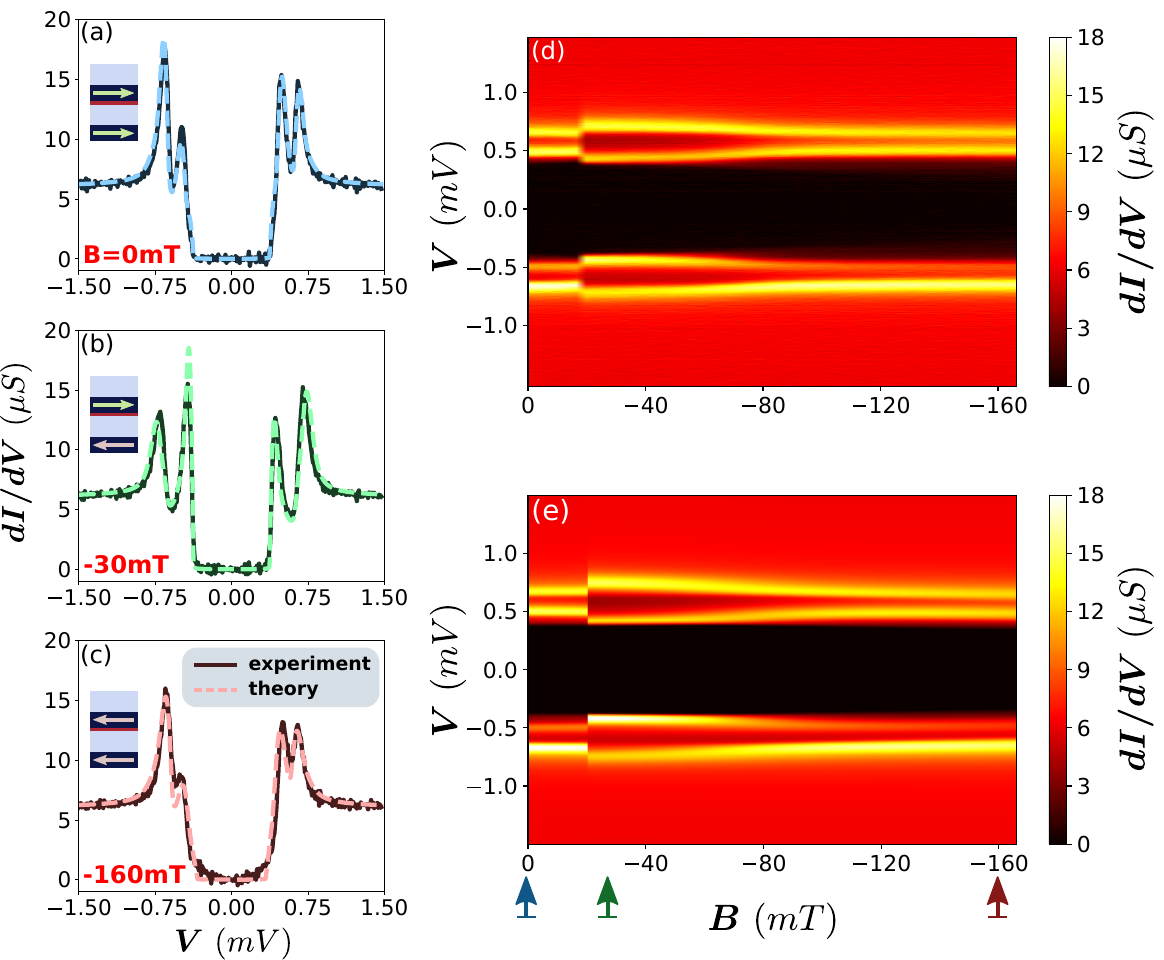}
  \caption{Magnetic-field dependence of the tunneling conductance of the
    spin-polarized junction. Before the measurement, the system is polarized
    with a positive magnetic field ($B = 160$mT). The differential conductance
    is then measured at different values of magnetic fields from 0 to
    $- 160 $mT. (a), (b) and (c) show three different curves measured at 0, -20
    and -160 mT, respectively. (d) shows the full measured B-dependence. Panel
    (e) is the fitting resulting from the theoretical model.  }
  \label{fig:Exp2}
\end{figure}

In this section we present our measurements of the current-voltage ($I$-$V$)
characteristic of a EuS(4)/Al(4)/AlO$_x$/EuS(1.2)/Al(4.3)~\footnote{During
  growth, the oxidation of the aluminum layer was not controled. Therefore, it
  does not necessarily have the stoichometry of Al$_2$O$_3$.} junction
(thickness in nanometers), see inset in Fig.~\ref{fig:Exp1}a. The samples
consist of cross bars fabricated by electron-beam evaporation on an in situ
metallic shadow mask with a typical junction area of 290$\times$290 $\mu$m$^2$.
\cite{strambini-2017-revealing}

The tunneling spectroscopy is obtained by measuring the $I\textrm{-}V$
characteristic in a DC two-wire setup, as sketched in the inset of
Fig.~\ref{fig:Exp1}a. From this measurement we determine the differential
conductance, $dI/dV$, via numerical differentiation. The measurements are done
at cryogenic temperatures in a filtered cryogen-free dilution refrigerator. We
first cool down the sample from room temperature to $25\,\textrm{mK}$ in a
non-magnetic environment. Before applying any external magnetic field, we
measure the $I$-$V$ characteristic (Fig.~\ref{fig:Exp1}a) and extract the
$dI/dV$ shown by the solid line in Fig.~\ref{fig:Exp1}b.  We then apply an
in-plane magnetic field (up to $160$ mT) strong enough to align the
magnetization of both EuS layers, and start decreasing it. During this process,
we measure the $I$-$V$ characteristic and determine the tunneling conductance at
each value of the applied magnetic field.  The full dependence is shown in the
color plot of Fig.~\ref{fig:Exp2}d. Panels (a-c) in Fig.~\ref{fig:Exp2}
correspond to different vertical cuts of Fig.~\ref{fig:Exp2}d at the positions
indicated by the arrows placed at the bottom of the figure.

The obtained tunneling conductance clearly shows the four-peak structure
expected from the spin-split superconducting density of states
(DOS)\cite{Hao:1990}. Notice that these peaks are also observed before applying
any magnetic field, Fig.~\ref{fig:Exp1}b. The position of the peaks in
Figs.~\ref{fig:Exp2}(a-c) is always symmetric with respect to the sign of the
applied voltage, however, after the first magnetization of the junction, their
heights are not. This behavior contrasts with the one shown in
Fig.~\ref{fig:Exp1}b for the demagnetized sample. The asymmetry is a fingerprint
of spin-polarized tunneling through the middle EuS thin
layer\cite{moodera-electron-1988,Hao:1990,meservey1994spin}, which only after
magnetization turns out to be apparent. In contrast, and according to the
physical picture provided in Sec.~\ref{sec:discussion}, when the sample is
demagnetized, the thin EuS barrier layer consists of magnetic domains smaller
than the coherence length with random polarization directions. This leads to a
negligibly small value of the induced spin-splitting field on the upper
superconductor and no spin-filtering effect on the current after averaging over
the junction area.

The separation between the peaks at positive (or negative) voltage,
Fig.~\ref{fig:Exp1}b and Fig.~\ref{fig:Exp2}(a-c), provides information about
the size of the spin-splitting energy induced in the Al layers. This splitting
is proportional to the effective exchange energy between the spins localized at
the EuS/Al interface and the Al conduction electrons\cite{zhang-2019-theory}.
 
We observe a sudden increase of the spin-splitting energy at -$20$ mT
(Fig.~\ref{fig:Exp2}d), which occurs when the system switches to the
antiparallel configuration. As it turns out from our theoretical discussion in
Sec.~\ref{sec:discussion}, it is the bottom EuS layer that switches first and
abruptly. By further increasing the magnetic field, $B$, the parallel
configuration is recovered gradually with a smooth switching of the middle EuS
magnetization. The two rather different switching behaviors of the EuS films can
be attributed to a different magnetic configuration and anisotropy of the two
films due to different deposition conditions, which crucially depends on the
growth morphology \cite{miao2009controlling,de2018toward}.

Whereas the peak positions can be explained by using a simple tunneling model
\cite{meservey1994spin}, detailed features such as the width and height of the
peaks can only be understood by taking into account different scattering and
depairing mechanisms and performing a self-consistent calculation of the
superconducting order parameter. With this aim, in the next sections we present
a theoretical model that allows us to describe the $dI/dV$ curves, extract the
values of the different parameters, and provide a physical picture that explains
the full behavior shown in Fig.~\ref{fig:Exp2}d.

\section{The Model}
\label{sec:model}

\begin{figure}[t!]
  \centering \includegraphics[width=.9\linewidth]{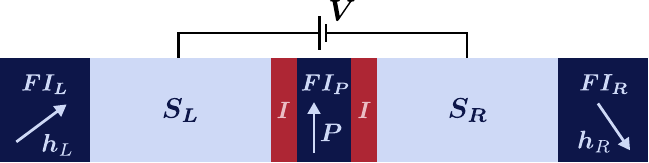}
  \caption{Schematic of a tunnel junction between two spin-split superconductors
    with a spin polarized tunneling barrier and biased at a voltage $V$. The
    left (right) superconductor $\rm{S_L}$ ($\rm{S_R}$) experiences a
    spin-splitting field $\boldsymbol{h}_L$ ($\boldsymbol{h}_R$) by an attached
    ferromagnetic insulator layer $\rm{FI_L}$ ($\rm{FI_R}$). The spin polarized
    tunneling barrier, with polarization $\boldsymbol{P}$, is another
    ferromagnetic insulator (FI). To avoid the magnetic proximity effect, the
    superconductors are separated from the spin-polarized tunneling barrier by
    insulating layers (I).  The superconductor $\rm{S_L}$ ($\rm{S_R}$) is at
    temperature $T_L$ ($T_R$).}
  \label{fig:sketch}
\end{figure}

In this section we present a theoretical model to describe the electronic
transport in junctions with spin-split superconductors and spin-filtering
barriers. The goal of this section is twofold: On the one hand, to obtain
general results for the current in tunnel junctions between two spin-split
superconductors in the presence of a spin-filtering barrier. On the other hand,
we provide a complete description of the experimental results presented in the
previous section.

We consider a generic junction, sketched in Fig.~\ref{fig:sketch}. It consists
of two spin-split superconductors separated by a spin-polarized tunneling
barrier. The spin-split superconductors correspond to two S/FI bilayers, whereas
the tunneling barrier is an additional FI layer with adjacent thin insulating
layers to decouple it magnetically from the superconductors.

To describe the current through the junction below, we use the tunneling
Hamiltonian approach, such that the system is described by
\begin{equation}
  \label{eq:generic-hamiltonian}
  H = H_L + H_R + H_T.
\end{equation}
Here $H_{L(R)}$ describes the left(right) superconducting electrode attached to
a FI and $H_T$ the tunneling of electrons between the
superconductors\cite{abrikosov-2017-fundamentals}.
  
In order to compute the current one needs to determine the spectral properties
of the decoupled FI-S electrodes.  We model them by assuming that the
interaction between the localized magnetic moments in the FI and the conduction
electrons in the S layer creates an effective exchange field in the
latter\cite{khusainov1996indirect,Kushainov_review,zhang-2019-theory,heikkila2019thermal}.
If the superconducting films are thinner than the coherence length such exchange
field can be assumed to be homogeneous in the S and, hence, the S electrodes are
described by
\begin{equation}
  \label{eq:H_electrodes}
  H_{L(R)}=H_{BCS}+{\bf h}_{L(R)}\cdot \boldsymbol{\hat\sigma}\; ,
\end{equation}
where ${\bf h}_{L(R)} = h_{L(R)}{\bf n}_{L(R)}$ is the exchange field pointing
in the direction of the unit vector ${\bf n}_{L(R)}$, $\boldsymbol{\hat \sigma}$
is the vector of Pauli matrices and $H_{BCS}$ is the BCS Hamiltonian that also
includes random impurities, magnetic, non-magnetic and those with spin-orbit
coupling\cite{maki1964-bst}.

For the tunneling Hamiltonian (the last term of
Eq.~\eqref{eq:generic-hamiltonian}) we assume that the tunneling through the
barrier is spin dependent; in other words, that the electron tunneling
probability depends on whether its spin is oriented parallel or anti-parallel 
with respect to the barrier magnetization\cite{Bergeret2012}.

We consider a general case where the directions of the magnetization in each of
the three FIs are independent of each other. A voltage $V$ is applied across the
junction and, in principle, the temperatures of the two FI/S electrodes are
different $T_L\neq T_R$. Here, the indices $L$ and $R$ denote the left and right
electrode respectively.

The effective splitting of the left and right superconductors in
Fig.~\ref{fig:sketch} is given by the induced exchange fields
$\boldsymbol{h}_L = h_L \boldsymbol{n}_L$ and
$\boldsymbol{h}_R = h_R \boldsymbol{n}_R$ respectively, whereas the spin
filtering is described by the polarization vector
$\boldsymbol{P} = P \boldsymbol{n}_P$ with
$P \equiv \frac{G_\uparrow - G_\downarrow}{G_\uparrow + G_\downarrow}$ and
b$0 \leq P \leq 1$. The vectors $\boldsymbol{n}$ are unit vectors pointing in
the respective directions, the magnitude of the exchange fields $h_{L/R}$ has
energy units and $G_{\uparrow(\downarrow)}$ stands for the tunneling conductance
through the junction for carriers with up (down) spin along the direction of
$\mathbf{n}_P$.

Without loss of generality, we set the barrier magnetization along the $z$ axis,
$\boldsymbol{n}_P = (0,0,1)$, such that the magnetization orientations of the
adjacent S/FI bilayers can be parametrized by three angles, $\theta_{L,R}$ and
$\gamma$:
\begin{equation}
  \label{eq:h-orientation-left}
  \boldsymbol{n}_L = (\sin\theta_L, 0, \cos\theta_L)
\end{equation}
and
\begin{equation}
  \label{eq:h-orientation-right}
  \boldsymbol{n}_R = (\sin\theta_R\cos\gamma, \sin\theta_R\sin\gamma,
  \cos\theta_R).
\end{equation}
In a collinear configuration, {\it i.e.} $\theta_L=\theta_R=0$, the current
through the junction can be straightforwardly calculated from the well-known
tunneling expression\cite{meservey1994spin}. We next generalize the latter for
non-collinear magnetizations. Moreover, in order to include the effects of spin
relaxation and depairing, we use the quasiclassical Green's functions (GFs) for
an accurate description of the spectrum of the S/FI electrodes.

\subsection{Quasiclassical Green's functions for spin-split superconductors}
\label{sec:GF}

In this section we present the quasiclassical Green's functions and the
expression for the current as a function of the applied voltage and temperature
bias (see also \cite{heikkila2019thermal}) for an arbitrary magnetic
configuration of the junction shown in Fig.~\ref{fig:sketch}. We restrict our
analysis to the tunneling limit, which corresponds to the experimental situation
when a FI is used as a barrier. In such case, one can treat each FI/S electrode
in Fig.~\ref{fig:sketch} independently.  In other words, we can calculate the
GFs, $\breve g_L(\epsilon)$ and $\breve g_R(\epsilon)$, for each
electrode. Moreover, one can first consider the case in which $V=0$ and
$\varphi=0$, where $\varphi$ is the phase difference between the
superconductors. Finite $\varphi$ and $V$ can then be added as gauge factors.

We use the Green's functions defined in the Keldysh$\otimes$Nambu$\otimes$spin
space ~\footnote{To simplify the notation we skip throughout the text the direct
  product symbol $\otimes$}.  These are are 8$\times$8 matrices that satisfy the
normalization condition
\begin{equation}
  \label{eq:GF-normalization}
  \breve g_s^2 = 1\; . 
\end{equation}
In the Keldysh space they can be written as\cite{Langenberg1986}:
\begin{equation}
  \label{eq:GF-keldysh-structure}
  \breve g_s = \left(
    \begin{array}{cc}
      \check g_s^R & \check g_s^K \\
      0 & \check g_s^A
    \end{array}\right),
\end{equation}
where $s=\{L,R\}$ labels left and right sides of the junction, $\check g_s^R$
stands for the retarded component of the GFs,
$\check g_s^A = -\hat\tau_3 \check g_S^{R^\dagger} \hat\tau_3$ is the advanced
component, and due to the normalization condition, the Keldysh component can be
written as
\begin{equation}
  \label{eq:keldysh-component}
  \check g_s^K = \check g_s^R \check f_s - \check f_s \check g_s^A.
\end{equation}
In these expressions, the "checks" $\breve\cdot\,$ indicate the full $8\times 8$
matrices, whereas $\check\cdot\,$ are used for $4\times 4$ matrices in
Nambu-spin space, and $\hat \cdot\,$ for $2\times 2$ matrices. $\hat\tau_i$ is
the $i$-th Pauli matrix in Nambu space and $\hat f_s$ stands for the electron
distribution function in electrode $s$.  In equilibrium, the latter is
proportional to the unit matrix in Nambu and spin space and reads:
\begin{equation}
  \label{eq:distribution-function-equilibrium}
  \check{f}_s(\epsilon) \equiv f_0(\epsilon,T_s) = \tanh \frac{\epsilon}{2k_BT_s},
\end{equation}
where $k_B$ is the Boltzmann's constant and $T_s$ is the temperature on the $s$
side of the junction. In our notation, whenever we do not specify any matrix
structure via Pauli matrices, it is implied that the matrix is proportional to
the unit matrix in the corresponding space.

We now calculate the GFs in the electrodes, which we assume in thermal
equilibrium. In the diffusive limit, they obey the Usadel
equation\cite{usadel.1970} with a local spin-splitting pointing in $z$-direction
and, as it was indicated after Eq.~\eqref{eq:GF-keldysh-structure} and in
Eq.~\eqref{eq:keldysh-component}, we only need to compute their retarded
component. Then, to calculate the current through the junction with
non-collinear magnetizations we will have to transform the GFs by using
spin-rotation operators.

The Usadel equation for the retarded component of a homogeneous S/FI electrode
in the spin local frame reads: \begin{equation}
  \label{eq:usadel-equation}
  \big[i\epsilon\hat\tau_3 - ih_s\hat\tau_3\hat\sigma_3 - \Delta_s \hat\tau_1 -
  \check \Sigma_s,\ \check g_s^R\big] = 0,
\end{equation}
where $\hat\sigma_i$ is the i-th Pauli matrix in the spin space and $\Delta_s$
is the self-consistent superconducting order parameter (see
Appendix~\ref{sec:app-self-consistency} for details).  The self energy,
$\check \Sigma_s$, consists of three contributions:
\begin{equation}
  \label{eq:self-energy-total}
  \check \Sigma_s = \check \Sigma_s^{so} + \check \Sigma_s^{sf} + \check\Sigma_s^{orb},
\end{equation}
the spin relaxation due to spin-orbit coupling, $\check \Sigma_s^{so}$, and
spin-flip relaxation, $\check \Sigma_s^{sf}$, and the orbital depairing,
$\check \Sigma_s^{orb}$, due to the external magnetic fields. Explicitly, each
contribution within the relaxation time approximation, reads:
\begin{align}
  \label{eq:self-energy-so}
  &\check \Sigma_s^{so} =
    \frac{\hat{\boldsymbol{\sigma}} \cdot \check g_s^R \cdot
    \hat{\boldsymbol{\sigma}}}{8 \tau_s^{so}}, \\[.5em]
  \label{eq:self-energy-sf}
  &\check \Sigma_s^{sf} =
    \frac{\hat{\boldsymbol{\sigma}} \cdot \hat\tau_3\check g_s^R \hat\tau_3 \cdot
    \hat{\boldsymbol{\sigma}}}{8 \tau_s^{sf}}, \\[.5em]
  \label{eq:self-energy-orb}
  &\check \Sigma_s^{orb} =
    \frac{\hat\tau_3\check g_s^R \hat\tau_3}{\tau_s^{orb}},
\end{align}
where $\tau_s^{so}$, $\tau_s^{sf}$ and $\tau_s^{orb}$ stand for spin-orbit,
spin-flip and orbital depairing relaxation times, respectively, and we use the
notation
$\hat{\boldsymbol{\sigma}}\cdot \check A \cdot \hat{\boldsymbol{\sigma}} =
\sum_{i=1}^3 \hat\sigma_i \check A \hat\sigma_i$ .

The general solution of the Usadel equation \eqref{eq:usadel-equation}, is then
given by four components in the Nambu-Spin space:
\begin{equation}
  \label{eq:GF-general-parametrization}
  \check g_s^R = (F_{0s} + F_{3s}\hat\sigma_3)\hat\tau_1
  + (G_{0s} + G_{3s}\hat\sigma_3)\hat\tau_3.
\end{equation}
The components proportional to $\tau_3$ are the normal components. They
determine the quasiparticle spectrum and enter the expression for the
quasiparticle current. The off-diagonal terms in Nambu space, here proportional
to $\tau_1$, are the anomalous GFs and describe the superconducting
condensate. They determine the Josephson current through the junction of
Fig. \ref{fig:sketch}. The anomalous GFs have two components: $F_{0s}$ describes
the singlet condensate, whereas the component $F_{3s}$ describes the triplet
component with zero total spin projection. Because we are considering diffusive
systems, both components have s-wave symmetry. This implies that the triplet
component is odd in frequency\cite{bergeret2001long}.  In
Sec.~\ref{sec:discussion} we numerically solve the Usadel equation,
Eq.~\eqref{eq:usadel-equation}, together with the normalization condition,
Eq.~\eqref{eq:GF-normalization}, and the self-consistent expression for
$\Delta_s$ (see Appendix~\ref{sec:app-self-consistency}).

We next derive the expression for the tunneling current in terms of the above
GFs.

\subsection{Tunneling current}
\label{sec:electric-current}

In the previous section we present the quasiclassical GFs,
$\breve g_s(\epsilon)$, in a local reference frame where $V=0$, $\varphi=0$ and
the exchange field is parallel to the $z$ axis.  We now use these results to
calculate the total electric current across the Josephson junction, sketched in
Fig.~\ref{fig:sketch}, in the presence of a finite voltage and phase difference,
and a non-collinear magnetic configuration.  This can be done by a gauge
transformation and a spin-rotation of the GFs.

In the presence of a voltage, the phase of a superconductor evolves in time as
\begin{equation}
  \label{eq:varphi-dynamic}
  \varphi(t) = \varphi + \frac{2eV}{\hbar}t\; ,
\end{equation}
where $\varphi$ is the dc phase. We define the corresponding gauge matrix
\begin{equation}
  \label{eq:GF-gauge-transformation}
  \hat U (t) = \exp \big(-i\varphi(t) \hat \tau_3\big)\; .
\end{equation}

If we assume that the voltage is applied on the left superconductor and the
magnetizations of the two S/FI and the spin-filter barrier are non-collinear
[see Eqs. (\ref{eq:h-orientation-left}-\ref{eq:h-orientation-right})] we can
obtain the GFs $\breve{\tilde{g}}$ from those obtained in in Sec~\ref{sec:GF}
via the following transformations:
\begin{align}
  \label{eq:GF-transformation-L}
  & \breve{\tilde g}_L(t-t') = \hat R_L \; \hat U(t) \;
    \breve g_L(t-t') \; \hat U(t')^\dagger \; \hat R_L^\dagger, \\
  \label{eq:GF-transformation-R}
  & \breve{\tilde g}_R(t-t') = \hat R_R \; \breve g_R(t-t') \; \hat R_R^\dagger. 
\end{align}
Here, the operators $\hat R_s$ describe spin-rotations in the left and right
electrodes:
\begin{align}
  \label{eq:GF-spin-rotation-L}
  &\hat R_L= \exp \big(-i \theta_L \hat\sigma_y / 2\big), \\
  \label{eq:GF-spin-rotation-R}
  &\hat R_R= \exp \big(-i \gamma \hat\sigma_z / 2\big)
    \exp \big(-i \theta_R \hat\sigma_y / 2\big)\; , 
\end{align}
and the time-dependent Green's functions in
Eqs.~(\ref{eq:GF-spin-rotation-L}-\ref{eq:GF-spin-rotation-R}) are obtained from
the GFs in frequency space:
\begin{equation}
  \label{eq:GF-fourier}
  \breve g_s(t-t') = \frac{1}{2\pi} \int_{-\infty}^\infty d\epsilon \;
  \breve g_s(\epsilon) \; e^{i\epsilon (t-t)}\; . 
\end{equation}

From Eqs.~\eqref{eq:GF-transformation-L} and \eqref{eq:GF-transformation-R} we
can now write the full expression for the time-dependent electric current across
the junction shown in Fig.~\ref{fig:sketch}:\cite{bergeret2012spin}
\begin{equation}
  \label{eq:current-general}
  I_c (t) = \frac{G_T \pi}{16 e} \ \text{Tr} \bigg( \hat\tau_3
  \Big[\breve{\tilde g}_L
  \ \overset{\circ}{,}\ 
  \check\Gamma \breve{\tilde g}_R\check\Gamma \Big]^K
  \bigg),
\end{equation}
where $G_T$ is the normal state conductance of the junction,
$[\cdot\, \overset{\circ}{,} \, \cdot]$ is a commutator of convolutions
\footnote{ When the operators depend only on the difference of times the
  convolution is defined as
 $$(A \circ B)(t) = \int_{-\infty}^\infty dt' A(t-t') B(t'-t).$$ Consequently,
 the commutator reads
 $$\Big[ A\,; B \Big] = (A\circ B)(t) - (B\circ A)(t).$$ }, the
superscript $K$ stands for the Keldysh component of the commutator and Tr stands
for the trace over the Nambu$\times$spin spaces.

Equation \eqref{eq:current-general} is valid in the tunneling limit. The matrix
$\check\Gamma$ describes the effect of the spin-filtering layer and is defined
as
\begin{equation}
  \label{eq:gamma-parametrizaton}
  \check\Gamma = u + v\hat\sigma_3\hat\tau_3\; , 
\end{equation}
where the parameters $u$ and $v$ depend on the polarization of the barrier $P$
as follows:
\begin{equation}
  \label{eq:gamma-u}
  u = \sqrt{\frac{1 + \sqrt{1-P^2}}{2}},
\end{equation}
\begin{equation}
  \label{eq:gamma-v}
  v = \sqrt{\frac{1 - \sqrt{1-P^2}}{2}}.
\end{equation}
One can easily check from these expressions that $u^2+v^2 = 1$, $2uv = P$ and
$u^2-v^2=\sqrt{1-P^2}$.

After a lengthy but straightforward algebra we obtain from
Eq.~\eqref{eq:current-general} the charge current through the junction which can
be written as the sum of three components:
\begin{equation}
  \label{eq:current-split}
  I_c(t) = I + J_1 \sin\left(\varphi + \frac{2eVt}{\hbar}\right)
  + J_2\cos\left(\varphi + \frac{2eVt}{\hbar} \right)\; . 
\end{equation}

Here $I$ is the quasiparticle tunneling current and the remaining part is the
Josephson current. Specifically, $J_1$ is the usual Josephson critical
current. The third term is proportional to the cosine of $\varphi(t)$. In a
non-magnetic Josephson junction this term is finite only at non-zero bias. In
the literature it is known as the $\cos\varphi$ term and has been widely
studied\cite{harris1974cosine,larkin1967-teb,barone1982physics}. Interestingly,
in a magnetic junction this term can be non-zero even when $V=0$. In this case
this term leads to the so-called anomalous Josephson current that appears in
certain magnetic system with spin-orbit coupling or inhomogeneous magnetization
\cite{buzdin08direct,zazunov2009,brunetti2013,Reynoso2008,Nesterov2016,konschelle2015theory,bergeret2015theory,Bobkov_magneto,yokoyama2014anomalous,silaev2017theta,silaev2017anomalous}
and is discussed in more detail in Sec. \ref{sec:Josephson}.

From Eq.~(\ref{eq:current-general}) we derive the expressions for the three
components of the current in terms of the GFs.  For the quasiparticle tunneling
current, first term in Eq.~\eqref{eq:current-split}, we obtain
\begin{widetext}
  \begin{align}
    &I = \frac{G_T}{2e} \int_{-\infty}^\infty
      d\epsilon \bigg[f_0(\epsilon+eV,T_L) - f_0(\epsilon,T_R)\bigg] \bigg\{
      P \Big[\mathcal{N}_{0L}(\epsilon+eV)
      \mathcal{N}_{3R}(\epsilon) \boldsymbol{n}_R\cdot\boldsymbol{n}_P + \mathcal{N}_{3L}(\epsilon+eV)
      \mathcal{N}_{0R}(\epsilon) \boldsymbol{n}_L\cdot\boldsymbol{n}_P \Big] \nonumber \\
    &\qquad\qquad\qquad + \mathcal{N}_{0L}(\epsilon+eV) \mathcal{N}_{0R}(\epsilon) + \mathcal{N}_{3L}(\epsilon+eV) \mathcal{N}_{3R}(\epsilon)
      \Big[\boldsymbol{n}_L^\parallel \cdot \boldsymbol{n}_R^\parallel + \sqrt{1-P^2} \ 
      \boldsymbol{n}_L^\perp\cdot\boldsymbol{n}_R^\perp \Big]
      \bigg\}, 
      \label{eq:current-I}
  \end{align}
\end{widetext}
where $\mathcal{N}_{is}(\epsilon) \equiv \text{Re}\big[G_{is}(\epsilon)\big]$ is
the semi-sum ($i=0$) and semi-difference ($i=3$) of the spin-up/spin-down
densities of states (DOS). In deriving this expression we have used the vector
equalities presented in Appendix~\ref{sec:app-unit-vectors}.

For the second and third terms in Eq.~\eqref{eq:current-split} we obtain
\begin{align}
  J_1 = &A_0\sqrt{1-P^2} + A_3\Big[
          \sqrt{1-P^2}\; \boldsymbol{n}^\parallel_L \cdot
          \boldsymbol{n}^\parallel_R \nonumber\\
        & + \boldsymbol{n}^\perp_L\cdot\boldsymbol{n}^\perp_R \Big] -
          B_3 \; P \boldsymbol{n}_P\cdot(\boldsymbol{n}_L\times \boldsymbol{n}_R)
          \label{eq:current-J1}
\end{align}
and
\begin{align}
  J_2 = &B_0\sqrt{1-P^2} + B_3\Big[
          \sqrt{1-P^2}\; \boldsymbol{n}^\parallel_L \cdot
          \boldsymbol{n}^\parallel_R \nonumber\\
        & + \boldsymbol{n}^\perp_L\cdot\boldsymbol{n}^\perp_R \Big] +
          A_3 \; P \boldsymbol{n}_P\cdot(\boldsymbol{n}_L\times \boldsymbol{n}_R),
          \label{eq:current-J2}
\end{align}
where $A_i$ and $B_i$ ($i=1,3$) are expressed in terms of the real and imaginary
part of the anomalous GFs $F_{is}(\epsilon)$:
\begin{widetext}
  \begin{align}
    & A_i = \frac{G_T}{2e} \int_{-\infty}^\infty d\epsilon \Big[
      f_0(\epsilon, T_R) \text{Re}\big[F_{iL}(\epsilon+eV)\big]
      \text{Im}\big[F_{iR}(\epsilon)\big] + 
      f_0(\epsilon+eV,T_L) \text{Im}\big[F_{iL}(\epsilon+eV)\big]
      \text{Re}\big[F_{iR}(\epsilon)\big] \Big],\label{eq:coefssA}
    \\[1em]
    & B_i = \frac{G_T}{2e} \int_{-\infty}^\infty d\epsilon \Big[
      f_0(\epsilon+eV, T_L) - f_0(\epsilon, T_R) \Big] \text{Im}
      \big[F_{iL}(\epsilon + eV)\big] \text{Im} \big[F_{iR}(\epsilon)\big].\label{eq:coefssB}
  \end{align}
\end{widetext}

Equations (\ref{eq:current-split}-\ref{eq:coefssB}) determine the total current
through the junction and are used in the next sections.  We start by analyzing
the Josephson current in magnetic junctions.

\section{Anomalous Josephson current}
\label{sec:Josephson}

An interesting situation occurs when $V=0$, $\varphi=0$, $T_L=T_R$ and the
magnetization vectors of the three FI layers are not in the same plane.  In this
case $I=0$, $B_0=B_3=0$ and the only term contributing to the current $J_2$ is
the one proportional to $A_3$ in Eq.~(\ref{eq:current-J2}). The latter is finite
when $\boldsymbol{n}_P\cdot(\boldsymbol{n}_L\times \boldsymbol{n}_R)\neq0$, {\it
  i.e.}, when three vectors are not co-planar. In this case a finite Josephson
current may flow through the junction even if the dc phase difference $\varphi $
is zero. This is the so-called anomalous Josephson current and the junction is
referred as a $\varphi_0$-junction. The latter has been widely studied in
magnetic junctions with spin-orbit coupling
\cite{buzdin08direct,liu2010relation,malshukov2010inverse,yokoyama2014anomalous,bergeret2015theory,konschelle2015theory,lu2019proximity}
or multilayer metallic ferromagnets
\cite{margaris2010zero,braude2007fully,Moor2015,Grein2009,liu2010,Mironov2015,silaev2017anomalous,silaev2017theta}.

In this section we discuss the possible observation of the anomalous Josephson
junction in FI/S-based junctions.  This effect was not yet seen in the samples
discussed here, because the large value of the normal-state resistance made it
impossible to measure any Josephson current at the temprature of the
experiments. However, similar type of samples with increased junction
transparency would be good candidates for measuring the $\phi_0$ effect.

Because we assume a unique temperature, $T_L=T_R=T$, and the junction is in
equilibrium ($V=0$), quasiparticle current is zero and one can write the
expression for the Josephson current in terms of a sum over Matsubara
frequencies. The anomalous functions proportional to the Pauli matrix $\sigma_3$
correspond to the odd-in-frequency triplet components of the condensate,
$F_3(i\omega_n)=-F_3(-i\omega_n)$, whereas those proportional to $\sigma_0$
arise from the singlet components $F_0(i\omega_n)=F_0(-i\omega_n)$
\cite{Bergeret2005}. The total current, Eq.~\eqref{eq:current-split}, can then
be written as
\begin{align}
  J_1 = &\ \pi T \frac{\pi G_T}{2e} \sum_\omega \bigg[\sqrt{1-P^2} \left(F_0^2 + F_3^2 \boldsymbol{n}^\parallel_L \cdot \boldsymbol{n}^\parallel_R\right) \nonumber\\
        & \qquad \qquad \qquad+ F_3^2\boldsymbol{n}^\perp_L \cdot \boldsymbol{n}^\perp_R \bigg]\\
  J_2 = &\ \pi T \frac{\pi G_T}{2e} P \boldsymbol{n}_P\cdot(\boldsymbol{n}_L\times \boldsymbol{n}_R) \sum_\omega F_3^2.
          \label{eq:anom} 
\end{align}
The contribution proportional to $\sin\varphi$ contains the conventional singlet
Josephson current that vanishes when the barrier is fully polarized $P=1$. If
the magnetizations and the barrier magnetization are non-collinear, there is an
additional contribution stemming entirely from the interference of the triplet
component of the condensate, as discussed in
Refs.~\onlinecite{Bergeret2012,bergeret2012spin}.

The anomalous current in Eq.~\eqref{eq:anom} is also a pure triplet current
which requires non-coplanar vectors, {\it i.e.} a finite triple product
$\boldsymbol{n}_P\cdot(\boldsymbol{n}_L\times \boldsymbol{n}_R)$, and it is
proportional to the polarization of the barrier. The well-defined splitting and
strong barrier polarization make the EuS/Al material combination suitable for
the realization of such magnetic anomalous junctions.

In the limit $T \rightarrow 0$ we obtain analytic results for the Josephson
current by assuming equal amplitudes of the exchange fields, $h_L=h_R \equiv h$,
and neglecting all relaxation processes,
$\tau_{so}^{-1} = \tau_{sf}^{-1} = \tau_{orb}^{-1} = 0$:
\begin{align}
  &J_1 = \frac{\pi G_T \Delta}{2e} \bigg[\sqrt{1-P^2}\eta + \nonumber \\
  &\qquad + \Big(\sqrt{1-P^2} \boldsymbol{n}^\parallel_L\cdot\boldsymbol{n}^\parallel_R + \boldsymbol{n}^\perp_L\cdot\boldsymbol{n}^\perp_R\Big)\Big(\eta-1\Big) \bigg],\label{eq:J1-anal}\\
  &J_2 = \frac{\pi G_T \Delta}{2e} P \big( \eta-1 \big) \boldsymbol{n}_P\cdot(\boldsymbol{n}_L\times \boldsymbol{n}_R),
\end{align}
where $\Delta$ is the real self-consistent superconducting order parameter at
zero temperature and exchange field $h$ and
\begin{equation}
  \eta \equiv \frac{32\Delta^2 (256\Delta^4 - 32\Delta^2h^2 + 9h^4)}{(16\Delta^2 - h^2)^3} - 1.
\end{equation}

In the case where $h=0$ (and, therefore,
$\boldsymbol{n}_L = \boldsymbol{n}_R = 0$), the coefficient $\eta=1$ and
Eq.~\eqref{eq:J1-anal} yields the well-known
Ambegaokar-Baratoff\cite{ambegaokar1963tunneling} formula for the Josephson
current with a prefactor $\sqrt{1-P^2}$ due to the barrier polarization.

\section{Quasiparticle current and differential conductance}
\label{sec:discussion}

In this section we discuss the quasiparticle current, Eq. (\ref{eq:current-I}),
and use our theoretical framework to describe the experimental data shown in
Figs.~\ref{fig:Exp1}-\ref{fig:Exp2}.  In the following discussion, we identify
the layer at the bottom (top) in the experimental setup, Fig.~\ref{fig:Exp1},
with the left (right) electrode of the model in Fig.~\ref{fig:sketch}.
 
The experimental setup corresponds to a situation in which the EuS barrier
serves two purposes: on the one hand, it acts as a spin-filtering barrier and,
on the other hand, it causes the spin-splitting in one of the superconductors
(the right one in Fig. \ref{fig:sketch}). This means that the orientation of
barrier magnetization coincides with the direction of the exchange field in the
right superconductor, $\boldsymbol{n}_P = \boldsymbol{n}_R$, while the
magnetization $\boldsymbol{n}_L$ is, in principle, independent of the
magnetization of the barrier. The left superconductor (S$_L$) is in a good
contact with the outer EuS, which induces a finite $h_L$.  At the other
interface between S$_L$ and the tunneling barrier, a thin oxide layer is formed,
preventing the exchange coupling \cite{Hao:1990}. Thus, for our specific sample,
the thinnest FI layer in the middle is a tunneling barrier (1.2 nm) which
induces the spin splitting only on the right superconductor and polarizes the
current, whereas the thicker EuS layer (4 nm) causes the spin splitting in the
left Al film.

Because the two EuS layers are of different thicknesses and they were grown on
two different substrates, it is expected that the magnetization switching is
different, as well as the strength of the induced exchange splittings in the
superconductors, $h_R\neq h_L$. We assume the same superconducting order
parameter, spin orbit and spin flip relaxation times for both Al films. Moreover
the temperatures are assumed to be equal, $T_L = T_R=T$.

Because of the high normal-state resistance of the tunneling barrier ($\sim160$
k$\Omega$), no Josephson current through the junction could be measured, as
shown in the left panel of Fig.~\ref{fig:Exp1}. In particular, the Josephson
energy $E_J = \hbar J_1/(2e)$ was of the order of the temperature and,
therefore, the thermal fluctuations of the phase smeared out the Josephson
effect.  The current shown in that figure corresponds only to the quasiparticle
contribution and it can be determined from Eq.~\eqref{eq:current-I} for
$\boldsymbol{n}_R\cdot\boldsymbol{n}_P = 1$ and $\boldsymbol{n}_R^\perp = 0$. We
can parametrize the magnetic configuration of the junction by a single angle
$\theta$ between the splitting field in the left and right superconductor:
$\boldsymbol{n}_R\cdot\boldsymbol{n}_L = \boldsymbol{n}_P\cdot \boldsymbol{n}_L
= \cos\theta$.

\begin{figure}[t!]
  \centering \includegraphics[width=\linewidth]{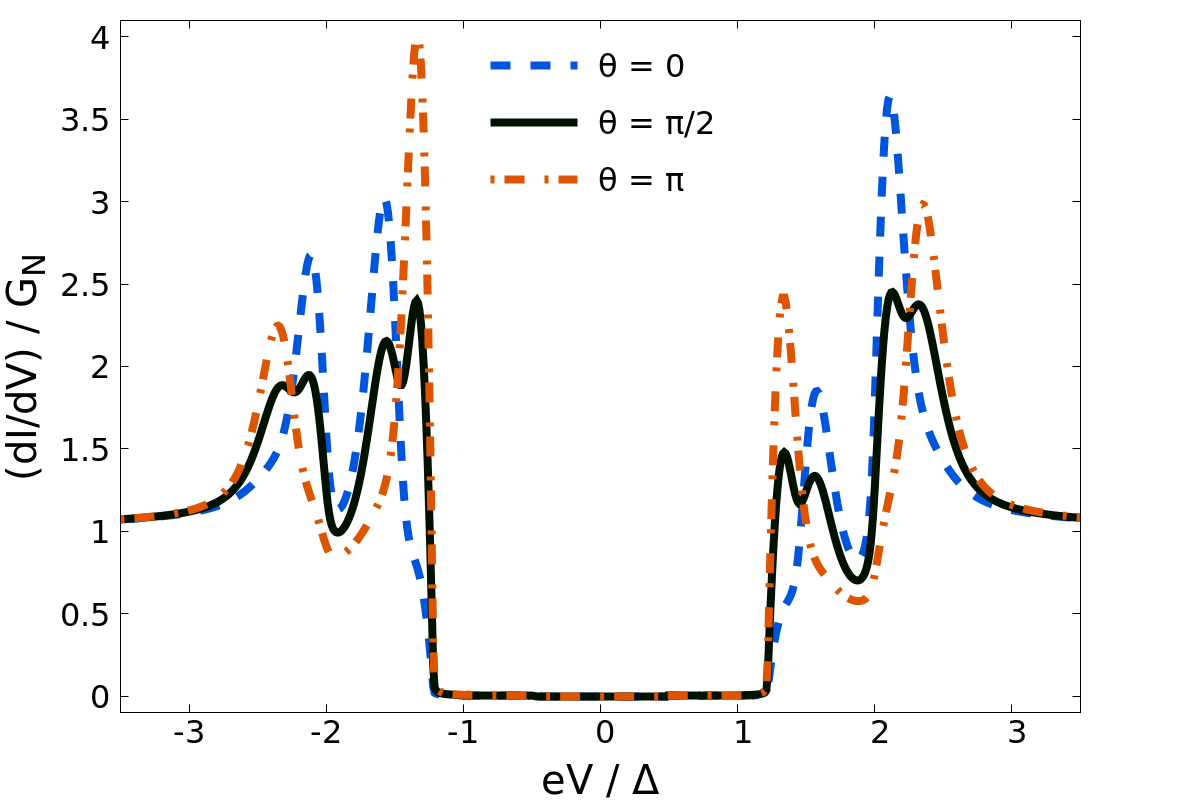}
  \caption{Normalized differential conductance spectrum of the
    FI$_L$/S$_L$/I/FI$_R$/S$_R$ junction calculated from our theoretical
    model. Both superconductors are assumed to have the same order parameter,
    $\Delta_0$. The polarization of the barrier is parallel to the exchange
    field induced in the right superconductor,
    $\boldsymbol{n}_P\parallel \boldsymbol{n}_R$, while the exchange field of
    the left superconductor forms an angle $\theta$ with $\boldsymbol{n}_R$. The
    dashed lines correspond to collinear situations, (blue) $\theta=0$ and (red)
    $\theta=\pi$, while the solid black line corresponds to a non-collinear one,
    $\theta = \pi/2$. The remaining parameters used in the calculation are
    $\tau_{so}^{-1} = \tau_{orb}^{-1} = 0$ and $\tau_{sf}^{-1} = 0.08\Delta_0$
    for the relaxation times in both superconductors, Zeeman splitting values of
    $h_L=0.35\Delta_0$ and $h_R = 0.10\Delta_0$, a polarization of $P=0.25$ and
    a global temperature of $k_BT = 0.01\Delta_0$.}
  \label{fig:current-theta}
\end{figure}

From Eq.~\eqref{eq:current-I}, we compute the current and, after differentiation
with respect to $V$, we obtain the differential tunneling conductance
$dI/dV$. In Fig.~\ref{fig:current-theta} we show its dependence on the voltage
for different values of the angle $\theta$ and certain values of spin splitting
fields and spin relaxation times. For a collinear configuration of
magnetizations, $\cos\theta = \pm 1$, the differential conductance shows the
four-peak structure, observed in most of experiments on EuS/Al based structures
\cite{moodera-electron-1988,Hao:1990,tedrow-spin-1986,meservey1994spin,
  strambini-2017-revealing}. These peaks appear at voltages
$eV=\pm (\Delta_L+\Delta_R) \pm (h_L - \cos \theta h_R)$.
 
However, if the magnetizations of the FIs are non-collinear, we find a
qualitatively new result (see the solid black line in
Fig.~\ref{fig:current-theta}). Instead of four peaks, the differential
conductance shows eight peaks for any value of $\theta$ between $0$ and
$\pi$. These two different behaviors can be understood as follows: In the
collinear case, the spin component along the single direction of magnetization
is globally conserved and the two spin species tunnel independently. When the
polarization of the tunneling barrier is non-collinear with the magnetization of
one of the electrodes, tunneling does not conserve spin. The additional peaks in
the $dI/dV$ stem from the projection of the electron spin of one of the
electrodes onto the local spin basis in the other electrode. The peaks in
$dI/dV$ then appear at $eV=\pm (\Delta_L+\Delta_R) \pm (h_L \pm h_R)$

This unusual situation occurs when the induced exchange field, and hence the
magnetization of the EuS films, is spatially homogeneous, so that the eight-peak
structure of $dI/dV$ shown in Fig.~\ref{fig:current-theta} can only be observed
if the EuS are monodomain magnets with non-collinear magnetizations. In our
EuS/Al samples the situation is rather different. As discussed in
Ref.~\onlinecite{strambini-2017-revealing}, EuS films consist of an ensemble of
crystallites with intrinsic magnetization
\cite{tischer1973ferromagnetic}. Therefore, before applying any external
magnetic field, the magnetic configuration of the EuS layers consists of
randomly oriented magnetic domains. Typically the size of EuS/Al tunnel
junctions (here $\sim 290 \times 290$ $ \mu$m$^2$) is much larger than the size
of these domains and, therefore, the measured tunneling current is determined by
an average over the angle $\theta$,
$\braket{I}_{\theta} \equiv \int_0^{\pi} \frac{d\theta}{\pi} \; I$, which reads:
\begin{align}
  &\braket{I}_\theta = \frac{G_T}{2e}\int_{-\infty}^\infty d\epsilon
    \Big[f_0(\epsilon+eV, T_L) - f_0(\epsilon, T_R)\Big] 
    \nonumber\\ &\ 
                  \times \Big[\mathcal{N}_{0L}(\epsilon+eV)\mathcal{N}_{0R}(\epsilon)
                  + P\mathcal{N}_{0L}(\epsilon+eV)
                  \mathcal{N}_{3R}(\epsilon)\Big].
                  \label{eq:current-demagnetized}
\end{align}

We use this averaging procedure to fit the experimental data shown in
Fig.~\ref{fig:Exp1}, which corresponds to the situation before any magnetic
field has been applied. As discussed above, the finite spin-filtering
coefficient $P$ results in an asymmetry in the $dI/dV$ curve with respect to the
sign of $V$. However, Fig.~\ref{fig:Exp1} shows a quite symmetric curve. This
can be explained by assuming that the domain size in the upper thin EuS layer is
smaller than $\xi_0$ and, therefore, the possible splitting in the corresponding
superconductor (R in our case) averages out. The absence of a Zeeman field in
the right superconductor leads to an equal density of states for up and down
electrons and, hence, $\mathcal{N}_{3R}(\epsilon) = 0$. Consequently, the second
term on the second line of Eq.~\eqref{eq:current-demagnetized} does not
contribute to the current, which now does not depend on the spin polarization of
the tunneling barrier.

The theory curve in Fig.~\ref{fig:Exp1}b (blue line), is obtained for $G_T = 6$
$\mu S$, which is the value of the conductance measured at sufficiently large
voltages (see the right panel of Fig.~\ref{fig:Exp1}). The superconducting gap
at zero field and zero temperature is set to $\Delta_0 = 320$ $\mu$eV in both Al
layers. According to previous studies on the spin relaxation processes in
aluminum layers \cite{jedema2002electrical,poli2008spin,bergeret2018colloquium},
we set the spin-orbit relaxation time to $\tau_{so}^{-1} = 0.005\Delta_0$. The
spin-flip relaxation is however enhanced due to the magnetic disorder caused by
the adjacent EuS layer and we chose $\tau_{sf}^{-1} = 0.08\Delta_0$ in both Al
layers.  Since the measurements in Fig.~\ref{fig:Exp1} are for zero field then
$\tau_{orb}^{-1} = 0$. The best fitting is obtained for $h_L = 100$ $\mu$eV
(bottom layer in the experiment), whereas $h_R = 0$ as explained above. The EuS
at the bottom is a thicker film and its magnetic domain size is of the order of,
or even larger than, the superconducting coherence length
$\xi_0$\cite{strambini-2017-revealing}. Therefore it induces a sizable exchange
splitting in the bottom Al layer.

We now focus on the results of Fig.~\ref{fig:Exp2} when an external field is
applied. These measurements are done after the first magnetization of EuS, {\it
  i.e.}, after a strong enough in-plane magnetic field is applied ($B=160$ mT).
After this, we switched off the $B$-field and measured the $I$-$V$
characteristic varying the magnetic field from $B=0$ to $B\approx-160$ mT. The
differential conductance obtained by a numerical differentiation is shown with
solid lines in panels (a-c) of Fig.~\ref{fig:Exp2} for $B=0$, $B=-30$ mT and
$B=-160$ mT, respectively. A full overview of the $dI/dV$ is presented as a
color map in panel Fig.~\ref{fig:Exp2}(d).

From the four-peak structure of $dI/dV$ and the theoretical prediction in
Fig.~\ref{fig:current-theta}, we can conclude that the average induced exchange
fields in the left and right superconductors are collinear. After the
application of the initial strong magnetic field, the magnetizations of both EuS
are aligned in the direction of $B$. By decreasing the field until it switches
its direction, the magnetization of the FIs may also switch at their
corresponding coercive fields leading to the usual ferromagnetic hysteresis
loop.  Such switching events can be seen from the evolution of the peak
positions in the $dI/dV$ map in Fig.~\ref{fig:Exp2}d.

We calculate the current using Eq.~\eqref{eq:current-I} and fit the data shown
in Fig. \ref{fig:Exp2}.  We use for the values of the spin-splitting fields for
large magnetic fields (saturation of the magnetization of the EuS films)
$h_{L}^{sat} = 120$ $\mu$eV and $h_{R}^{sat} = 30$ $\mu$eV. The difference
between the values of the exchange fields after and before the first
magnetization of the EuS layers is consistent with the result in
Ref.~\onlinecite{strambini-2017-revealing}.  In order to describe the evolution
of the conductance peaks with the magnetic field we assume that the exchange
field follows the evolution of the local magnetization. In particular, for the
color plot in Fig.~\ref{fig:Exp2}e we assume that
$h_L(B) = h_{L}^{sat} \cdot y_L(B)$ and $h_R(B) = h_{R}^{sat} \cdot y_R(B)$,
whereas spin-polarization of the barrier is chosen to be
$P(B) = 0.25 \cdot y_R(B)$. Here, $y_L(B) = 1 - 2\theta (B + 20)$ and
$y_R(B) = \tanh \frac{B + 70}{40}$ are two empirical functions that describe the
evolution of the magnetization in the bottom and top EuS layers as a function of
the magnetic field $B$ given in mT, where $\theta(x)$ is the step function.

We also take into account the orbital depairing in the superconducting layers
due to the applied magnetic field, determined by
\cite{anthore-2003-density,deGennes:566105}
\begin{equation}
  \label{eq:tau-orb}
  \tau_{orb}^{-1} = \left(\frac{\pi d \xi_0 B}{\sqrt{6} \Phi_0}\right)^2 \Delta_0,
\end{equation}
where $\Phi_0$ is the magnetic flux quantum, $d\approx$ 4 nm is the width of the
Al layers and $\xi_0 \approx$ 200 nm is the superconducting coherence length.

The results of our fitting procedure are the dashed lines in panels (a), (b) and
(c) of Fig.~\ref{fig:Exp2} and the color map in panel (e). All in a good
agreement with the experimental data.

At first glance our fitting suggests an unexpected behavior: the thin EuS layer
switches its magnetization slower than the thicker one. Here we provide a
plausible explanation for this behavior, which can be caused by the different
polycrystalline structures of EuS layers grown under different conditions. The 4
nm thick EuS (bottom layer in Fig.~\ref{fig:Exp2}) is grown on an Al$_2$O$_3$
substrate, while the 1.2 nm barrier is grown directly on the previously oxidized
underlying Al layer. As the oxidation of this layer is not controlled, its
stoichiometry is completely different to the one on top of the substrate. Most
likely, the thin layer consists of a more disordered set of crystallites and
islands, resembling a superparamagnet. Such a large structural roughness could
also arise from the propagation of defects created during growth in the bottom
EuS and Al layers. If the RMS roughness is larger than half thickness of the top
EuS layer, the layer would become discontinuous.  Thus, the different
thicknesses of the two EuS layers plays an important role in determining their
magnetic properties as well. Presumably, the crystallites in the thick EuS layer
are magnetically well coupled, while in the thin magnetic layer they form
decoupled magnetic islands. Consequently, the EuS in the bottom would form
magnetic domains on a scale much larger than the crystallite size, which leads
to the sharp switching of the magnetization observed around $B=-20$ mT in
Fig.~\ref{fig:Exp2}d. In the thin EuS layer, by contrast, the macroscopic
magnetization is an average over the magnetization of the crystallites. Due to
disorder, the anisotropy is also random and such crystallites would not switch
simultaneously, resulting into the gradual magnetization reversal that we
observe from $B\approx-60$ mT to $B\approx-100$ mT in Fig.~\ref{fig:Exp2}d.
Moreover, the assumption of an island-like structure due to the growth
morphology\cite{miao2009controlling} can also explain the low polarization of
the FI layer (25\%) in comparison with previous results of near to 80\%
polarization\cite{tedrow-spin-1986,moodera-electron-1988}. Indeed, it seems that the
coverage of the EuS barrier is not complete and, in addition to the spin
polarized current, there is a parallel direct tunneling current through the
AlO$_x$ layer.

\section{Conclusions}
\label{sec:conclusions}

We present an exhaustive analysis of tunnel junctions between spin-split
superconductors coupled via a spin-polarized barrier. With the help of a
theoretical model, we compute the spectral properties of the S/FI electrodes and
determine the current through a FI/S/I/FI/I/S/FI junction, where the middle FI
layer serves as a spin-filter. Our theory predicts a previously unknown behavior
of the differential tunneling conductance when the FI layers are
non-collinear. Moreover, we suggest how to use these structures for the
realization of so-called $\varphi_0$-junctions.  In addition, our theory
provides an accurate description of the differential conductance measurements of
an EuS/Al/AlO$_x$/EuS/Al tunnel junction. We obtain diverse information from the
comparison between theory and experiment. On the one hand we can determine the
values for the induced spin-splitting fields, spin-filter efficiency, magnetic
disorder, spin-orbit coupling, and orbital effects in the superconductors. On
the other hand, from the magnetic field dependence of the $dI/dV(V)$ curves, we
can extract information about the magnetic structure of the two EuS layers,
which turns out to be very different due to the rather different growth
morphology of each layer.

\section*{Acknowledgments}
This work was supported by EU's Horizon 2020 research and innovation program
under Grant Agreement No. 800923 (SUPERTED).  MR, VNG and FSB, acknowledge
financial support by the Spanish Ministerio de Ciencia, Innovacion y
Universidades through the Projects No. FIS2014-55987-P and FIS2017-82804-P.
E. S. and F. G acknowledge partial financial support from the European Union's
Seventh Framework Programme (FP7/2007-2013)/ERC Grant 615187- COMANCHE, and by
the Tuscany Region under the FARFAS 2014 project SCIADRO. SC, FA and TTH
acknowledge support from the Academy of Finland (Key Funding project 305256 and
project number 317118).  The work of JSM at MIT was supported by NSF Grant
DMR-1700137, ONR Grant N00014-16-1-2657 and ARO grant W911NF1920041.
 
\appendix

\section{Self-consistency equation}
\label{sec:app-self-consistency}

The superconducting gap for each superconductor in the paper is obtained
self-consistently. In the quasiclassical theory, the self-consistency equation
is given by
\begin{equation}
  \Delta_s=\frac{\lambda}{16i}\int_{-\Omega_D}^{\Omega_D}d\epsilon\text{Tr}\left[\left(\tau_1-i\tau_2 \right)\check{g}^K_s(\epsilon) \right],
\end{equation}
where $s=\{L,R\}$ labels the superconductor, $\lambda$ is the coupling constant
and $\Omega_D$ is the Debye cutoff energy. Using the expression for the Keldysh
component in Eq.~\eqref{eq:keldysh-component} and the parametrization of the
Green's functions shown in Eq.~\eqref{eq:GF-general-parametrization}, we can
rewrite the self-consistency equation of the superconducting gap as
\begin{equation}
  \Delta_s=\frac{\lambda}{2}\int_{-\Omega_D}^{\Omega_D}d\epsilon\
  \text{Im}\left[F_{0s}(\epsilon)\right]\tanh\left(
    \frac{\epsilon}{2k_BT}\right)\; ,
\end{equation}
We use this self-consistent superconducting gap, together with the Usadel
equation in Eq.~\eqref{eq:usadel-equation} to calculate the Green's functions
used in current calculations.

\section{Relations between unit vectors}
\label{sec:app-unit-vectors}

In order to derive the expressions for the quasiparticle current and
supercurrents in Sec.~\ref{sec:model}, we made use of following relations
between the unit vectors pointing in the direction of the polarization of the
barrier, $\boldsymbol{n}_P$, and induced the exchange fields in the left,
$\boldsymbol{n}_L$, and right, $\boldsymbol{n}_R$, electrodes. We define the
parallel and perpendicular components of the exchange fields with respect to the
polarization vector:
\begin{align}
  &\boldsymbol{n}_s^\parallel \equiv (\boldsymbol{n}_s\cdot \boldsymbol{n}_P)
    \boldsymbol{n}_P = \cos\theta_s \boldsymbol{n}_P, \\
  &\boldsymbol{n}_s^\perp \equiv \boldsymbol{n}_s - \boldsymbol{n}_s^\parallel,
\end{align}
where $s=\{L,R\}$ labels the position of the electrode. According to these
definitions and the expressions for the unit vectors of the Zeeman fields in
Eqs.~\eqref{eq:h-orientation-left} and \eqref{eq:h-orientation-right}, we obtain
the following useful relations:
\begin{align}
  & \boldsymbol{n}_L \cdot \boldsymbol{n}_R = \boldsymbol{n}_L^\parallel \cdot
    \boldsymbol{n}_R^\parallel + \boldsymbol{n}_L^\perp \cdot
    \boldsymbol{n}_R^\perp, \\
  & \boldsymbol{n}_L^\parallel \cdot \boldsymbol{n}_R^\parallel = \cos\theta_L
    \cos\theta_R, \\
  & \boldsymbol{n}_L^\perp \cdot \boldsymbol{n}_R^\perp = \sin\theta_L
    \sin\theta_R \cos\gamma, \\
  & \boldsymbol{n}_P \cdot (\boldsymbol{n}_L \times \boldsymbol{n}_R) =
    \sin\theta_L \sin\theta_R \sin\gamma.
\end{align}



\bibliographystyle{apsrev4-1} \bibliography{list.bib}

\end{document}